\documentclass{appolb}%
\pdfoutput=1
\usepackage{graphicx}%
\usepackage{amsmath}%
\usepackage{amsfonts}%
\usepackage{amssymb}
\begin{document}

\title{Large-$N_{c}$ pole trajectories of the vector kaon $K^{\ast}(892) $ and of
the scalar kaons $K_{0}^{\ast}(800)$ and $K_{0}^{\ast}(1430)$%
.\thanks{Presented at XI Workshop on Particle Correlations and Femtoscopy, 3-7/11/2015, Warsaw.}
}
\author{Milena So{\l }tysiak$^1$, Thomas Wolkanowski$^2$ and \\Francesco Giacosa$^{1,2}$ 
\address{
$^1$Institute of Physics, Jan Kochanowski University, PL-25406 Kielce, Poland
$^2$Institute for Theoretical Physics, Goethe
University, D-60438 Frankfurt am Main, Germany}
}
\maketitle

\begin{abstract}
We study the spectral functions, the poles and their trajectories for increasing $N_{c}$ of the vector kaon state $K^{\ast}(892),$ characterized by
$I(J^{P})=\frac{1}{2}(1^{-})$, and of the scalar kaons $K_{0}^{\ast}(800)$ and $K_{0}^{\ast}(1430),$ characterized by 
$I(J^{P})=\frac{1}{2}(0^{+})$. 
To this end, we use relativistic QFT's Lagrangians with
both derivative and non-derivative terms. In the vector kaonic sector the
spectral function is well approximated by a Breit-Wigner function:
there is one single peak and, correspondingly, a single pole in the complex
plane. On the contrary, in the scalar sector, although the Lagrangian contains
only one scalar kaonic field, we find two poles, one corresponding to a
standard quark-antiquark ,,seed\textquotedblright\ state $K_{0}^{\ast}(1430),$
and one to a \textquotedblleft companion\textquotedblright\ dynamically generate
pole $K_{0}^{\ast}(800)$. The latter does not correspond to
any peak in the scalar kaonic spectral function, but only to an enhancement in
the low-energy regime.

\end{abstract}

%\eqsec  % uncomment this line to get equations numbered by (sec.num)

%\PACS{PACS numbers come here}

\section{Introduction}

Understanding the nature of the mesonic resonances listed in Ref. \cite{pdg} is an
important topic of both experimental and theoretical hadron physics. In the
vector kaonic sector with $I(J^{P})=\frac{1}{2}(1^{-})$ ($I$ stands for
isospin, $J$ for total spin, and $P$ for parity), the resonance $K^{\ast
}(892)$ corresponds very well to the expected quark-antiquark states $u\bar
{s},$ $d\bar{s},$ $s\bar{u}$, $s\bar{d}$; moreover, its spectral function is
nicely described by a (relativistic) Breit-Wigner function. On the contrary,
the scalar kaonic sector $I(J^{P})=\frac{1}{2}(0^{+})$ is much more
complicated. Two resonance are listed in the PDG below 1.5 GeV: the broad but
well-established $K_{0}^{\ast}(1430)$ and the very broad and light
$K_{0}^{\ast}(800)$ (also known as $\kappa$), whose existence still requires
definitive confirmation (for discussions, see e.g. Refs.
\cite{oller,dullemond,black,our} and refs. therein). The inclusion of $\kappa$
in the summary table of PDG would allow to complete the nonet of light scalar
states in the energy region below $1$ GeV.

The aim of this work, based on\ Ref. \cite{our}, is to study the nature of
the vector and scalar kaonic resonances. By using QFT Lagrangians, we determine the
coordinates of the poles on the complex plane and study their nature by using
the large-$N_{c}$ limit. We confirm that $K^{\ast}(892)$ and $K_{0}^{\ast
}(1430)$ are standard quark-antiquark states, while $K_{0}^{\ast}(800)$ is a
dynamically generated state.

\section{The model(s) and results}

In order to describe $K^{\ast}(892)$ and $K_{0}^{\ast}(1430)$ we introduce
relativistic Lagrangians that couples them to one kaon and one pion:%
\begin{equation}
{\mathcal{L}_{v}=cK^{\ast}(892)_{\mu}^{+}\partial^{\mu}K^{-}\pi^{0}+\ldots
}\text{ , }\mathcal{L}_{s}=aK_{0}^{\ast+}K^{-}\pi^{0}+bK_{0}^{\ast+}%
\partial_{\mu}K^{-}\partial^{\mu}\pi^{0}+\ldots.
\end{equation}
The expressions above contain non-derivative and derivative interaction terms. The
dots stay for the sum over isospin and Hermitian conjugation. The terms in\ Eq.
(1) naturally emerge as a subset of more complete mesonic models, e.g. Ref.
\cite{elsm}. According to our models, the decay widths of ${K^{\ast}(892)}$ and $K_{0}^{\ast}$ (as function of the running mass $m$) are: %
\begin{equation}
\Gamma_{{K^{\ast}}}(m)=3\frac{\left\vert \vec{k}_{1}\right\vert }{8\pi m^{2}%
}\frac{c^{2}}{3}\left[  -M_{\pi}^{2}+\frac{(m^{2}+M_{\pi}^{2}-M_{K}^{2})^{2}%
}{4m^{2}}\right]  F_{\Lambda}(m)\text{ ,}%
\end{equation}%
\begin{equation}
\Gamma_{K_{0}^{\ast}}(m)=3\frac{\left\vert \vec{k}_{1}\right\vert }{8\pi
m^{2}}\left[  a-b\frac{m^{2}-M_{K}^{2}-M_{\pi}^{2}}{2}\right]  ^{2}F_{\Lambda
}(m)\text{ ,}%
\end{equation}
where the form factor $F_{\Lambda}(m)=\exp(-2\vec{k}_{1}^{2}/\Lambda^{2})$ has
been introduced. $\Lambda$ is an energy scale, $\vec{k}_{1}$ the
three-momentum of one outgoing particle, $M_{K}$ the kaon mass, and $M_{\pi}$
the pion mass. The on-shell decay widths are obtained by setting $m$ to the
masses of ${K^{\ast}(892)}$ or $K_{0}^{\ast}(1430)$: this is accurate in the
former case, but quite imprecise in the latter. The (scalar part of the)
propagator of the resonances is given by $\Delta_{{K^{\ast}/}K_{0}^{\ast}%
}(p^{2}=m^{2})=\left[  m^{2}-M_{0}^{2}+\Pi(m^{2})+i\varepsilon\right]
^{-1}$, where $M_{0}$ is the bare mass of $K^{\ast}(892)/K_{0}^{\ast}(1430)$
and $\Pi(m^{2})=Re(m^{2})+iIm(m^{2})$ is the one-loop contribution. The
spectral function which determines the probability that resonance has a mass
between $m$ and $m+dm$ reads: $d_{{K^{\ast}/}K_{0}^{\ast}}(m)=\frac{2m}{\pi
}|{\operatorname{Im}}\Delta_{{K^{\ast}/}K_{0}^{\ast}}(p^{2}=m^{2})|.$ Spectral
functions must be normalized to unity. For details of the used formalism, see Ref.   \cite{thomas}.

The spectral functions of $K^{\ast}(892)$ and $K_{0}^{\ast}(1430)$ are shown
in\ Fig. 1. For the vector kaon we observe a single peak close to $0.9$ GeV
and a (unique!) pole at $0.89-i0.028$ GeV. In the scalar 
sector, there is a broad peak at about $1.4$ GeV, but no peak
corresponding to the light $\kappa$ (there is only a broad enhancement in the
low-energy regime). In this channel, it turns out that there are two poles: $(1.413\pm0.057)-i(0.127\pm0.011)$ GeV, which corresponds 
to the seed state $K_{0}^{\ast}(1430)$, and $(0.745\pm0.029)-i(0.263\pm0.027)$
GeV, which corresponds to $K_{0}^{\ast}(800)$ as an additional companion pole.
The parameters for the scalar channel were determined via a fit to existing
pion-kaon data, see Ref. \cite{our} for details.

We studied the change of the spectral function for different values of the number of colors
$N_{c}$ with the rescaling $a\rightarrow a\sqrt{3/N_{c}}$
(and so for $b$ and $c$). When $N_{c}$ increases, the interaction becomes smaller. In both channels we observe that the peak becomes narrower and
higher. However, in the scalar sector the enhancement of the $\kappa$ becomes smaller for increasing $N_{c}$. 

In Fig. 2 we show the trajectories of the poles for
increasing $N_{c}$. One sees that the
poles of $K^{\ast}(892)$ and $K_{0}^{\ast}(1430)$ tend to the real axis, while
that of $K_{0}^{\ast}(800)$ goes away from it and finally disappears for
$N_{c}\simeq13.$ In conclusion, all these results, with special focus on Fig. 2 which is the main outcome of the present proceedings, confirm that
$K_{0}^{\ast}(800)$ is a dynamically generated non-quarkonium meson.
\begin{figure}
\begin{center}
\includegraphics[width=0.4 \textwidth] {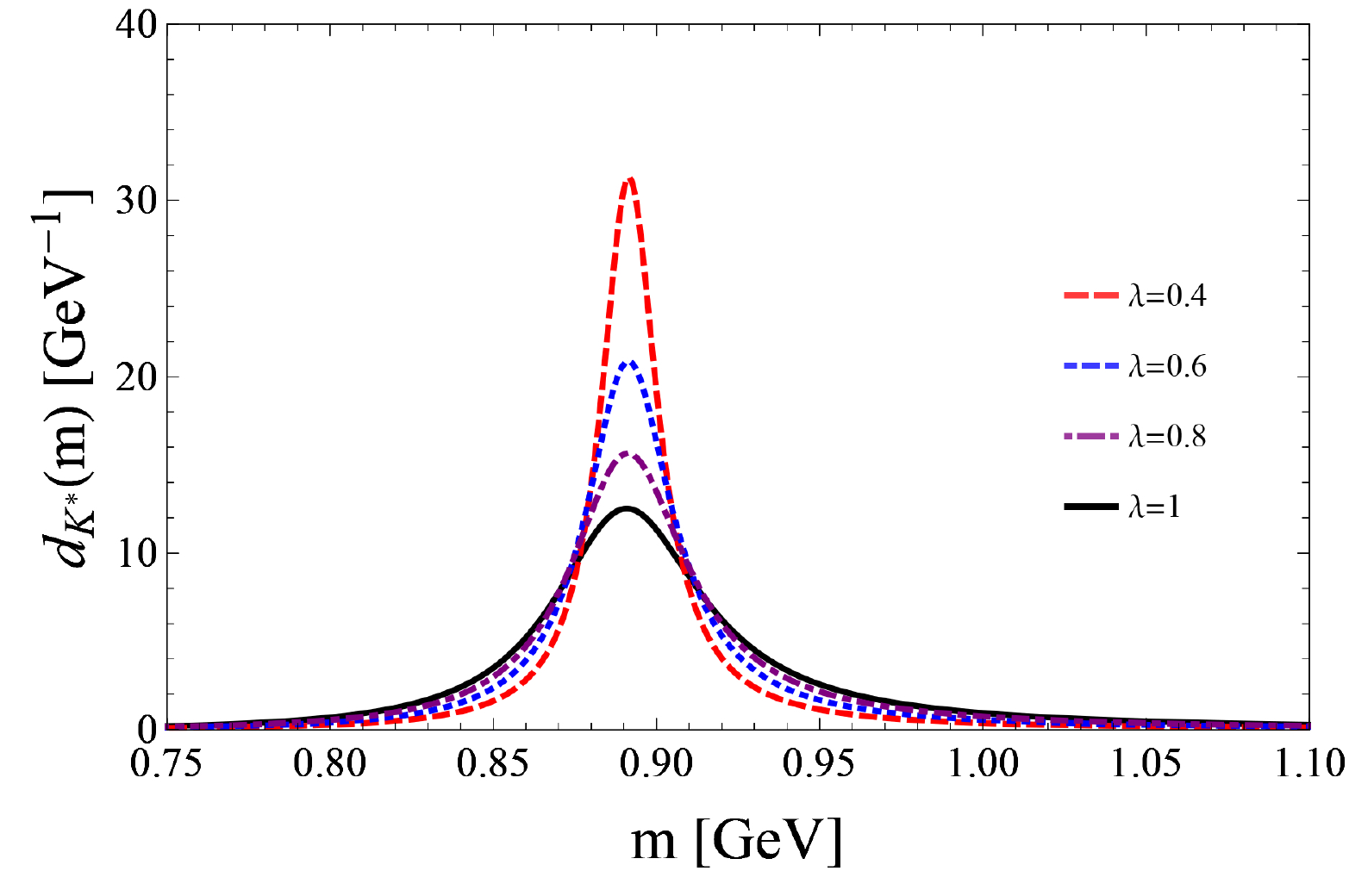}
\includegraphics[width=0.4 \textwidth] {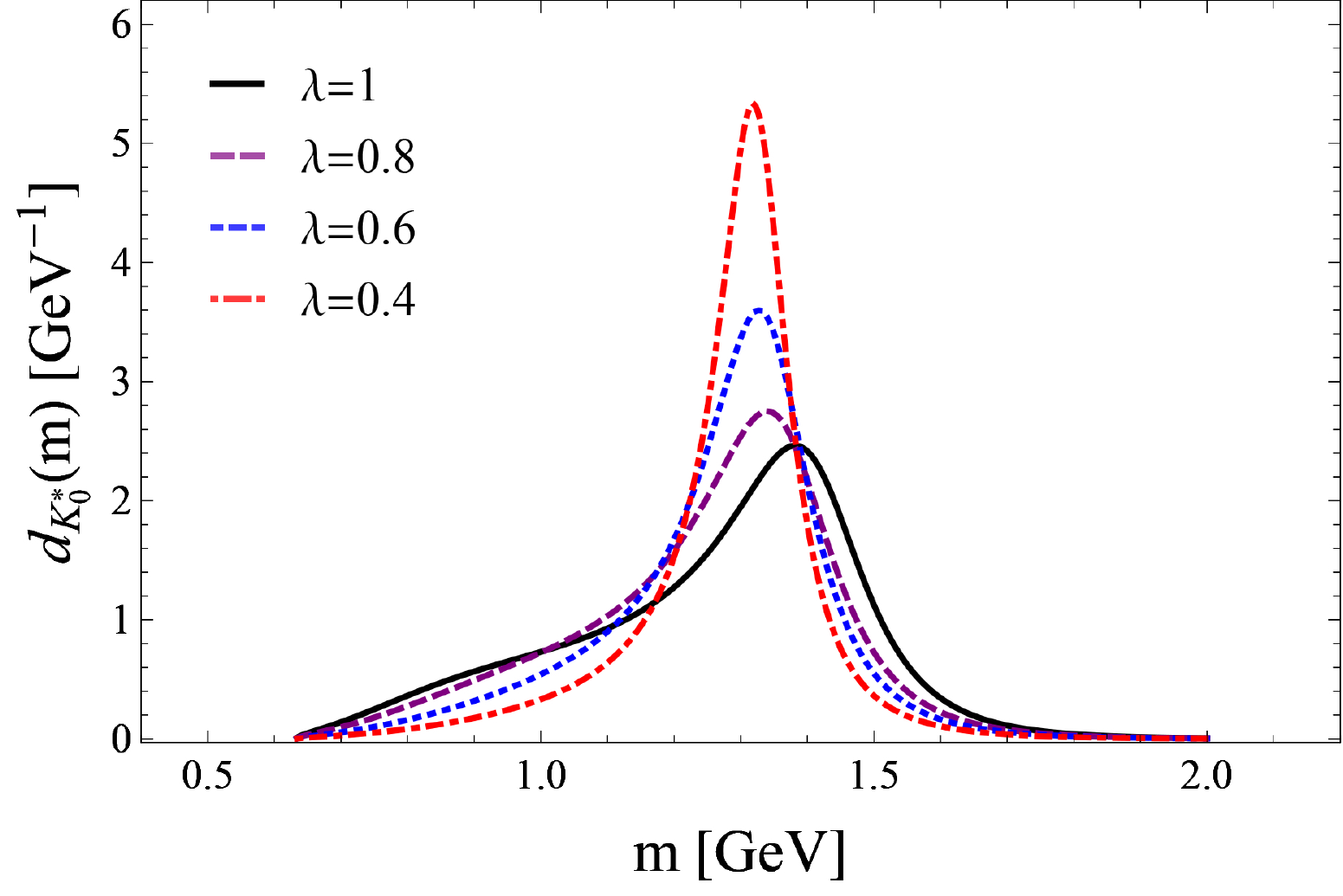}
\end{center}
\caption{Spectral functions of $K^{\ast}(892)$ (left
panel) and $K_{0}^{\ast}(1430)$ and of $K_{0}^{\ast}(800)$ (right panel) for different values of $\lambda=3/N_c.$
}%
\end{figure}

\begin{figure}
\begin{center}
\includegraphics[width=0.4 \textwidth] {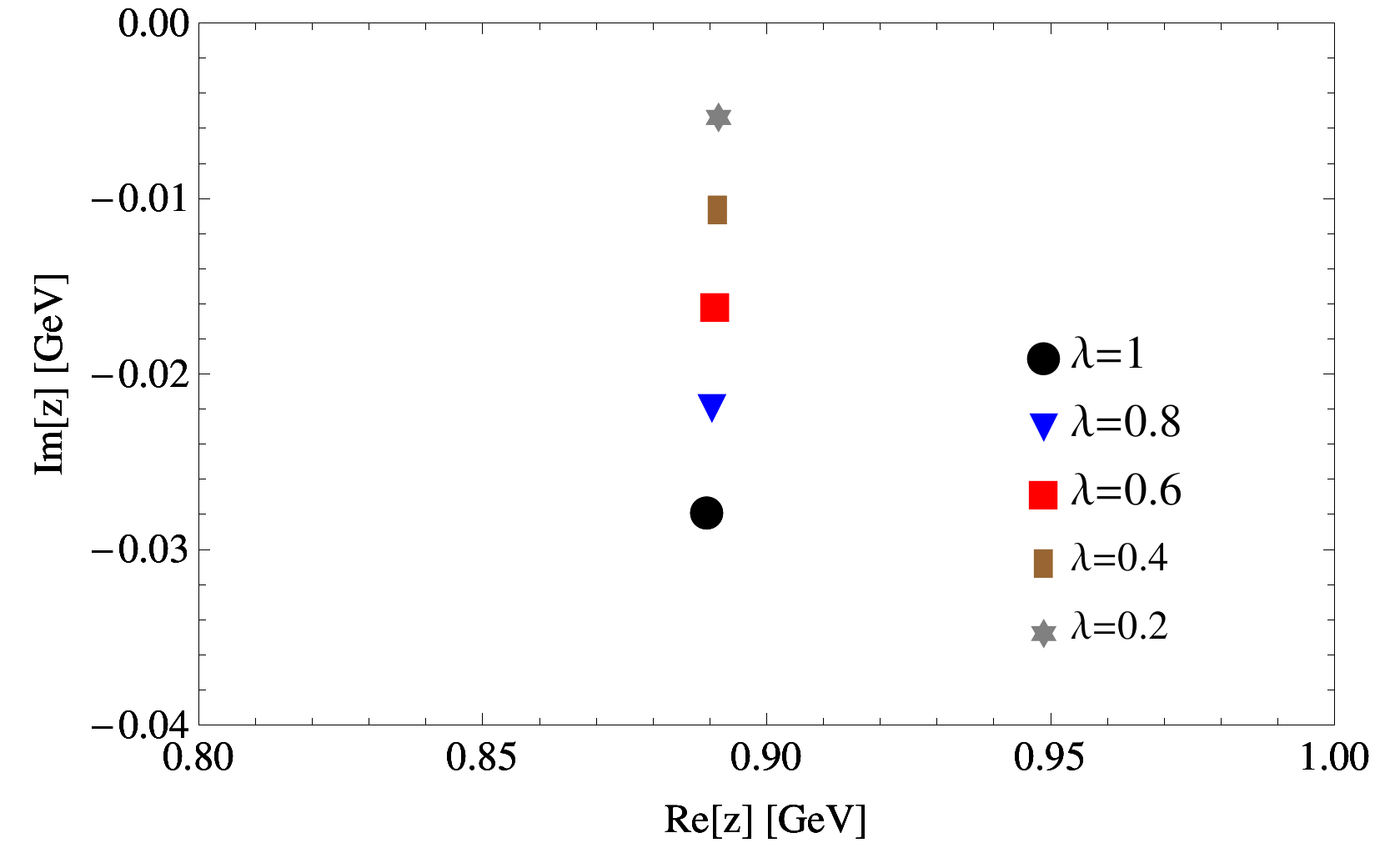}
\includegraphics[width=0.4 \textwidth] {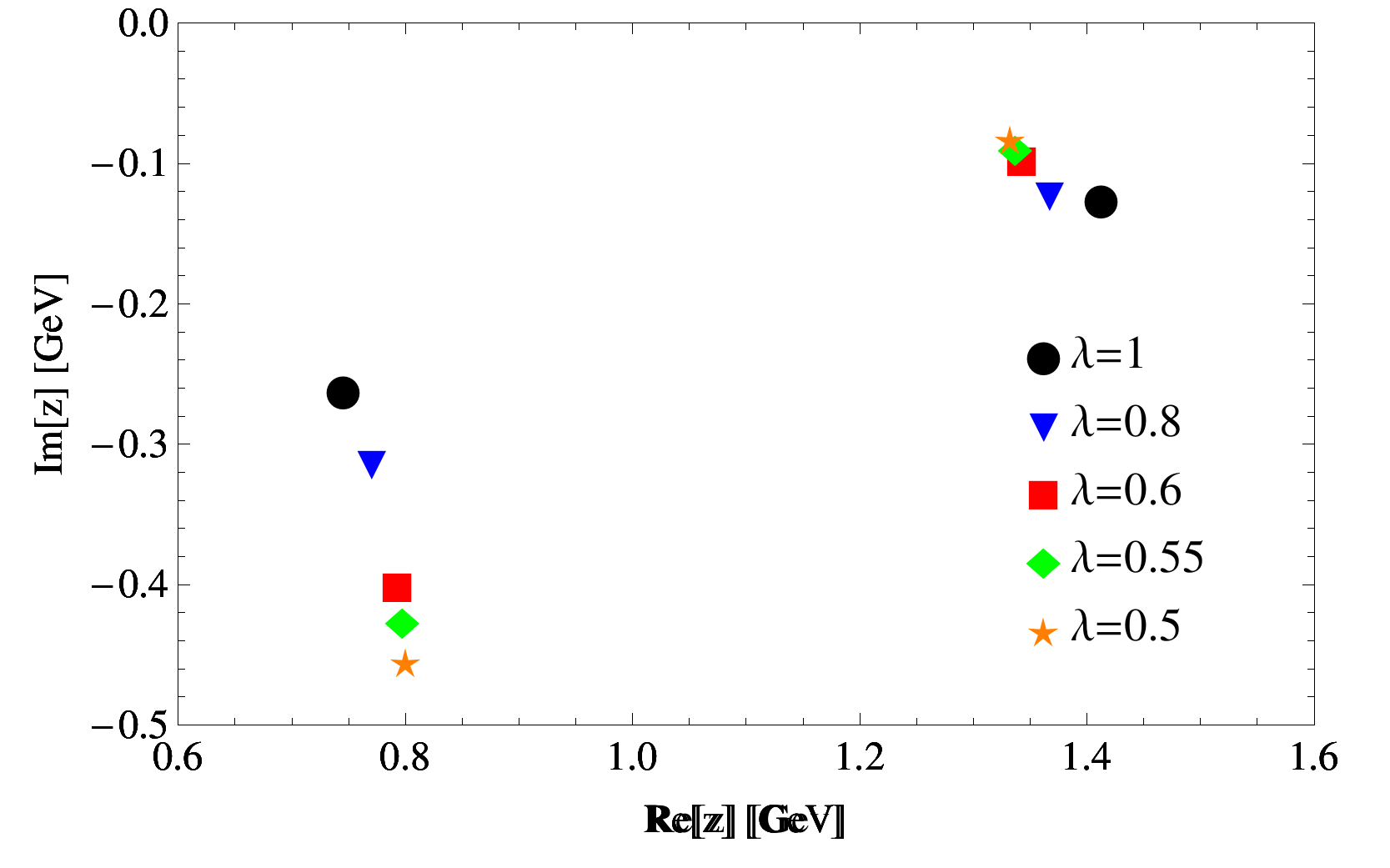}
\end{center}
\caption{Movement of the poles of $K^{\ast}(892)$ (left panel) and of $K_{0}^{\ast}(1430)$ and $K_{0}^{\ast}(800)$ (right panel) for
different values of $\lambda=3/N_c$. }%
\end{figure}

\newpage

\section{Conclusions}

We have discussed the nature of $K^{\ast}(892)$, 
$K_{0}^{\ast}(800)$, and $K_{0}^{\ast}(1430)$ by using QFT models presented
in\ Ref. \cite{our}. We find that $K^{\ast}(892)$ and $K_{0}^{\ast}(1430)$ are
regular quark-antiquark mesons (see the quark-model review in \cite{pdg}). In both cases, the spectral function  has a well-pronounced peak; in the large-$N_{c}$
limit the positions of the poles tend to the real axis, as it should for
conventional mesons. On the contrary,  $K_{0}^{\ast}(800)$ does not correspond
to a peak of the scalar spectral function, but there is a related pole in the
complex plane. The original Lagrangian contains a single scalar field, which
is associated to  $K_{0}^{\ast}(1430)$, hence $K_{0}^{\ast}(800)$ emerges as a
companion pole of $K_{0}^{\ast}(1430)$. In the large-$N_{c}$ limit the pole of
$K_{0}^{\ast}(800)$ disappears, confirming its non-quarkonium nature.

\end{document}